# Sequence Engineering of Copolymers using Evolutionary Computing

Ashwin A Bale[1] and Tarak K Patra[2*]

[1]Department of Chemical Engineering, Birla Institute of Technology and Science Pilani-Hyderabad, Hyderabad, TS 500078, India

[2]Department of Chemical Engineering, Center for Atomistic Modeling and Materials Design and Center for Carbon Capture Utilization and Storage, Indian Institute of Technology Madras, Chennai TN 600036, India

## Abstract

The correlations between the sequence of monomers in a macromolecule and its three-dimensional (3D) structure is a grand challenge in polymer science and biology. The properties and functions of macromolecules – both synthetic polymers and biomolecules depend on their 3D shape that has appeared to be dictated by their monomer sequence. However, the progress towards understanding the sequence-structure-property correlations and their utilization in materials engineering are slow because it is almost impossible to characterize astronomically large number of possible sequences of a copolymer using traditional experimental and simulation methods. To address this problem, here, we combine evolutionary computing and coarse-grained molecular dynamics (CGMD) simulation and study the sequence-structure correlations of a model AB type copolymer system. The CGMD based evolutionary algorithm screens the sequence space of the copolymer efficiently and identifies wide range of single molecule structures including extremal radius of gyrations ($R_gs$). The data provide new insights on the sequence-$R_g$ correlations of the copolymer system and their impact on the structure and functionality of polymeric materials. The work highlights the opportunities of sequence specific control of macromolecular structure for designing materials with exceptional properties.

# INTRODUCTION

Copolymers are a special class of macromolecules that are made of multiple chemical moieties with a rich phase behaviour[1–3] and tunability in their thermophysical properties.[4–9] Recent progresses in synthetic chemistry such as solid supported reaction, chemo-selective reaction,[10] orthogonal organic reaction,[11,12] acid-catalysed cascade reaction[13] and soluble phase synthetic strategy with molecular sieving[14] have been enabling to exercise an unprecedented level of control over sequences of different chemical moieties in a copolymer molecule.[15,16] These precision polymers provide enormous opportunities for molecular scale materials design with wide range of microstructure, functionality and activity being included in a polymeric material.[17–25] In a single molecular level, a copolymer's folding and three dimensional structure can be controlled to a great extent by adjusting its monomer sequences. Consequently, the bulk properties of polymeric materials such as elasticity, conductivity, rigidity, biodegradability, anti-bacterial activity and optoelectronic property can be precisely tuned by controlling the sequence of monomers in their constituting polymers. For example, copolymers with varying sequences are employed to improve the thermodynamic stability of polymer mixtures, and they, therefore, have wide applicability in emulsions and composite materials.[5,26,27] Even monomer level sequence specific control can lead to compatibilizers that yield zero interfacial energy between two immiscible homopolymers.[28,29] Similarly, the sequence of monomers in a charged copolymer is found to control its structural phase transitions,[30,31] liquid-liquid phase transition and associated critical point.[32] The morphology and interface of tapered block copolymers are found to be largely controlled by the details of the sequence of monomers in the middle block that is present in between two immiscible homogeneous blocks.[33–35] These works clearly suggest that minor changes in the sequence lead to a large variation in the structure, dynamics and phase behaviour of macromolecules. Moreover, the structure and properties of a copolymer in water and other solutions are also strongly influenced by its monomer sequence. In biological context, intrinsically disordered proteins (IDP) and structured proteins are primarily differed by their amino acid sequences. Why does a particular sequence of monomer prefer crystalline structure while a slight variation in the sequence yields disordered structure is not well understood. Among non-crystalline structures, how does a specific sequence determine swelling or collapsing of a polymer chain is also an unsolved problem. In spite of clear connections between sequence and three dimensional structure of a macromolecule, understanding and establishing their relationship is a complex and largely unsolved problem. The primary bottleneck to address this problem is the vast combinatorial sequence space of a macromolecule that precludes efficient screening. For instance, a linear copolymer chain with $n$ number of possible monomers and $m$ type of chemical moieties will have $m^n/2$ sequences that need to be explored, excluding the double counting due to the fact that a sequence and its reverse sequence represent one copolymer. Even for a AB type copolymer i.e. 2 types of chemical moieties with a chain length of 100, the total combinations possible are $2^{99}$ which is close to $10^{30}$. In practice, many copolymers possess more than two chemical moieties and several hundreds of monomer units. Given such an enormous sequence space that needs to be explored, it is highly desirable to minimize the number of property measurements – computer simulations or experiments needed to arrive at a sequence that corresponds to a desired target property and sample data across the entire space to establish sequence-structure correlations. Many of the previous studies have focused on diblock copolymers or copolymers with periodic blocks of monomers wherein a copolymer is usually characterized by its mean block length and mass fraction. The number of possible periodic sequences in a given copolymer of fixed chain length is very small and manageable using traditional experiment and simulation methods. However, the sequence space is mostly occupied by the non-periodic sequences that are astronomically large in number. Due to this vastness, non-periodic sequences are not well explored and whether mean block length can explain the variability of non-periodic copolymer's properties is not known. Therefore, the progress towards developing a comprehensive understanding of sequence-structure-property correlations is limited by the astronomically large number of possible non-periodic sequences of a copolymer that are impossible to characterize using traditional approaches. Towards this end an improved understanding of sequence-structure relations that are valid for all possible sequences (periodic and non-periodic) are essential for their successful adaptation in materials engineering. This requires rapid characterization and uniform sampling of the sequence space of copolymers.

To tackle this problem, machine learning and advanced optimization method are recently used for designing new polymer with target structures,[36] and characterizing sequence specific aggregates[37] in solution phase. Also, new copolymer sequences in bulk phase have been identified using evolutionary algorithm that improve thermal conductive of polymers.[38] Further, in our recent study, we have used coarse-grained molecular simulation based evolutionary search and Monte Carlo tree search to efficiently identify optimal sequences of polymer compatibilizers.[28,29] These optimization methods, particularly evolutionary methods, are progressively being used as a versatile strategy for material property optimization and design.[39] They have been increasingly adopted in the field of polymers[40–44] and other materials design problems[45–48] as well. Motivated by these previous studies, here, we employ evolutionary computing to understand the sequence space, conformational space of an 1:1 AB type linear copolymer and design target structures. Designing sequences for complex structures of copolymers are always challenging. It requires a priori understanding of the interactions of the building blocks and sequence-morphology correlations. For example, Khokholov and co-workers[8,49,50] introduce "instant image" strategy for protein-like homopolymer design. Within this strategy, a homopolymer with strong attractive interaction between all the monomers are simulated to form a globule structure. Then, monomers on the surface of the globule are identified by human intervention and hydrophilic interactions are selectively assigned to them. Unlike "instant image" approach, here our object is to develop a simple and autonomous method for target structure design. Moreover, we aim to assess the impact of sequence on a copolymer structure by generating homogeneous sequence-structure data across the entire sequence space. These two interconnected objectives are addressed by combining implicit solvent molecular dynamics (MD) simulation and evolutionary algorithm (EA). Our MD-EA determines the optimal sequences of a copolymer that give rise to its highest and lowest in equilibrium radius of gyration in an implicit solvent environment. The MD-EA workflow rapidly identifies target sequences of extremal gyrations despite the search space extending from one million to $10^{30}$. The MD-EA algorithm screens around 3000 candidate sequences to identify an optimal candidate and has proved to be very efficient and scalable for sequence engineering. The data suggest that both the compact structure and loose structure of the copolymer correspond to non-periodic sequences for all the polymer chain lengths studied in this work. The low Rg structures are similarly to globular protein-like aggregates that consist of a hydrophobic monomers in the core and polar groups on the surface. The current design scheme can identify protein-like copolymers autonomously. The data generated using this molecular simulation based evolutionary computing are analyzed to establish new sequence-structure correlations of copolymers. We further test the impact of sequence on the packing of a single macromolecule as well as bulk polymeric material. We demonstrate that the known range of material structures and assemblies can be expanded by monomer-to-monomer sequence control. The work outlines important design rules of sequence defined polymers, and has fundamental implications in designing exceptional polymeric materials.

## MODEL AND METHODS

**Model Copolymer.** A generic coarse-grained model of a copolymer is considered to study the impact of sequence on its structure. In this model system, two adjacent coarse-grained monomers of the copolymer are connected by the Finitely Extensible Nonlinear Elastic (FENE) potential of the form $E = -\frac{1}{2}KR_0^2 \ln\left[1 - \left(\frac{r}{R_0}\right)^2\right]$, where $K = 30\epsilon/\sigma^2$ and $R_0 = 1.5\sigma$. The pair interaction between any two monomer is modelled by the Lennard-Jones (LJ) potential of the form $V(r) = 4\varepsilon\left[\left(\frac{\sigma}{r}\right)^{12} - \left(\frac{\sigma}{r}\right)^6\right]$. The $\epsilon$ is the unit of pair interaction energy. The size of all the monomers is σ. The LJ interaction is truncated and shifted to zero at a cut-off distance $r_c = 2.5\sigma$ to represent attractive interaction among the monomers. The AA interaction strength is $\epsilon_{AA} = \epsilon$, while the BB and AB type interaction strengths are $\epsilon_{BB} = \epsilon_{AB} = 0.2\epsilon$. This ensures immiscibility between A and B type monomers. These set of model parameters are previous used to study copolymer systems,[51–53] and are very successful in understanding the generic properties of both synthetic polymers and biomolecules.

**Molecular Dynamics Simulation.** We conduct implicit solvent molecular dynamics simulations of a copolymer chain in a canonical ensemble. The initial configuration of a polymer chain is placed in a cubic simulation box of fixed dimension and box is periodic in all three directions. We use the Velocity Verlet algorithm with a timestep of $0.001\tau$ to integrate the equation of motion. Here, $\tau = \sigma\sqrt{m/\epsilon}$ is the unit of time, and $m$ is the mass of a monomer, which is same for both the A and B type moieties. All the simulations are conducted at a reduced temperature $T^* = T\epsilon/k_B = 1$, which is maintained by the Langevin thermostat within the LAMMS simulation environment.[54] The simulations are conducted for four different polymer chains of length N= 20, 50, 100 and 150. All the simulations are equilibrated for $10^7$ MD steps followed by a production run of $10^7$ steps. The data during the production cycles are collected for computing the equilibrium properties of the polymer chains. We further conduct simulations of bulk system for few selected sequence specific copolymers in a isothermal isobaric (NPT) ensemble. For NPT simulations, the temperature and pressure are maintained by the Langevin thermostat and Nose-Hoover barostat, respectively. All the bulk simulation are conducted at a reduced temperature *T\*=1*, and reduced pressure *P\*=0*. For each of these bulk simulations, 50 polymer chains are placed in a periodic simulation box, and the system is equilibrated for $10^7$ steps followed by an production run of $10^7$ steps.

**Molecular Simulation based Evolutionary Computing.** The central component of the sequence space screening and molecular design framework in this work is the use of an optimization algorithm for selection of candidate copolymer sequences en route to target properties. We adopt a metaheuristic approach viz. evolutionary computing for this purpose. The evolutionary computing is a class of algorithms that aim to mimic the process of natural selection to optimize a system's properties. These algorithms are commonly known as evolutionary algorithms (EAs) that begin with a set of initial candidates and compute an objective function quantifying their properties relative to target values. Afterword, the algorithms iteratively select new candidates in an effort to converge to a desired value of the objective function. The evolutionary processes usually continue until a criterion for convergence is satisfied or they are terminated because of a maximal time constraint. In these algorithms, new candidates are iteratively selected via several evolutionary operators – selection, crossover, mutation and elitism.[55,56] Here, we choose one such strategy that is suitable for the design of a binary copolymer with 1:1 composition of moieties and the objective function is its radius of gyration. The schematic structure of the molecular simulation based evolutionary algorithm that is employed for copolymer design and sequence space exploration is shown in Figure 1. It begins with a random population of 48 candidate copolymers. Each copolymer's monomer sequence is mapped to a binary genomic representation, as the model copolymer consists of only two types of chemical moieties. In the binary genome representation, 0 indicates A-type moiety and 1 indicates B-type moiety. Since, we consider equal proportion of A and B in the polymer chain, the number of 0s and 1s in the binary genome are equal in number. The fitness of all the candidates, which are their radius of gyration, is determined simultaneously by conducing multiple MD simulations in parallel. The MD simulations within the EA are conducted for sufficiently long time, and we ensure that the value of Rg converges over the simulation time. Then the top 16 out of the available 48 candidates are selected based on their ranking and moved to the next generation. These 16 candidates are the 'parents' who are used to produce new candidates using the crossover and mutation operations in a particular generation. In each step, two parents are randomly selected for crossover operation. The EA employs two-points crossover to combine the selected parents.[57] Afterward, a point mutation with a rate of 0.01 is applied to generate a new candidate. In the newly generated candidates, if the ratio of two genes deviates from 1:1, the EA applies additional point mutations on the gens such that the desired ratio is maintained. During the mutation, a randomly chosen gene changes its type from 0 to 1 or vice versa (bit flip) as shown in Figure 1. The gene selection for mutation are also random. This process continues until EA produce required number of new candidates in a generation. Therefore, every new generation would again consist of 48 new candidates: 16 of which are the parents retained from the previous generation and 32 new candidates generated using the evolutionary operators. The entire evolutionary cycle is iteratively continued for around 75 generations for all the cases. The total number of generation is decided based on the criteria that the optimal fitness does not improve over more than 10 generations.

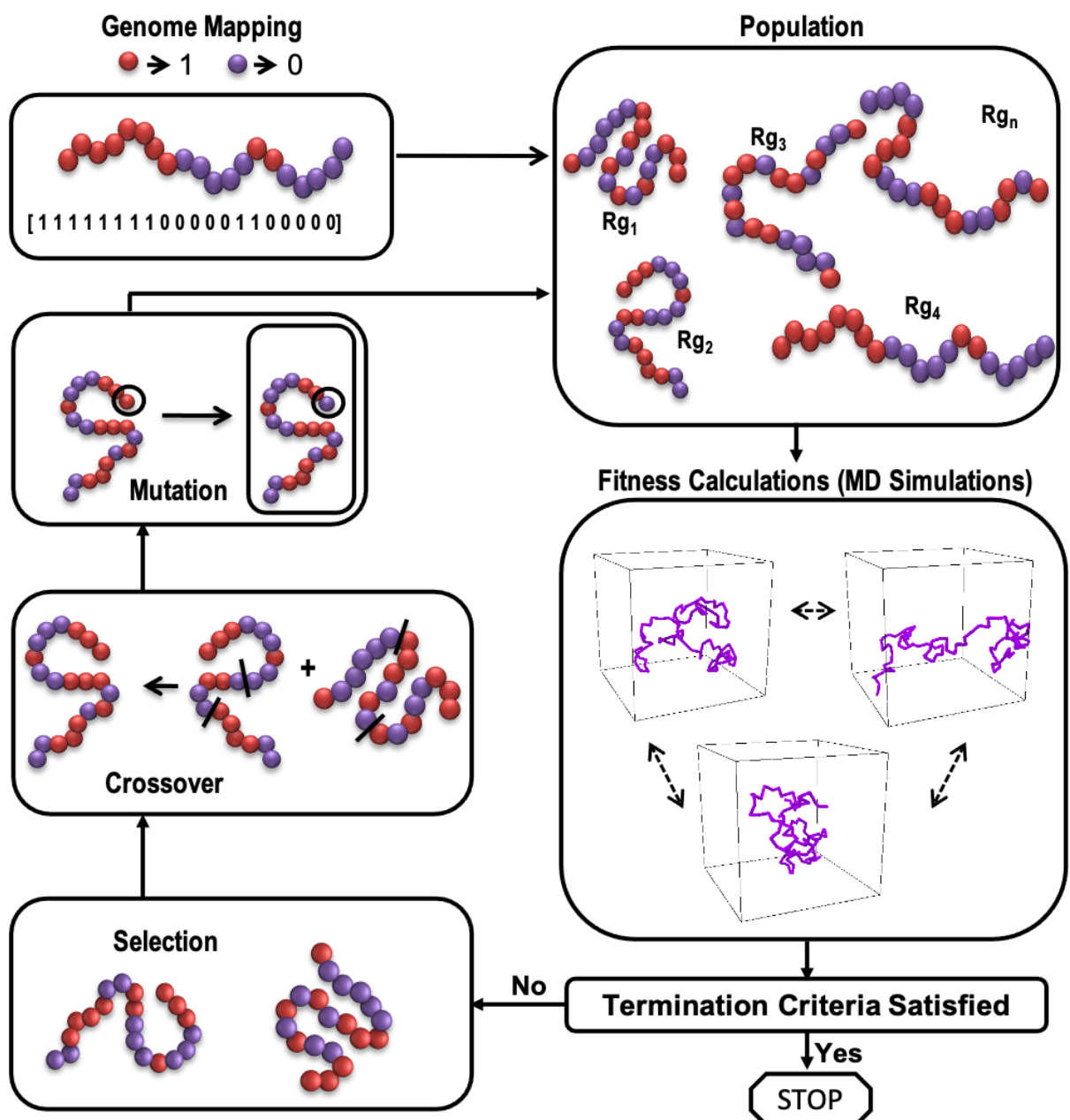

*Figure 1: Evolutionary computing for polymer design and sequence sampling. An AB type copolymer is mapped to a binary genome where A and B moieties are represented by 1 and 0, respectively. In each cycle, evolutionary operations – selection, crossover and mutation are applied to the binary sequence to create new copolymers. The fitness of copolymers i.e. the properties are computed using molecular dynamics simulations. Molecular simulations for all the candidates in a particular generation are conducted parallelly. For all the case studies, the evolutionary cycles are carried out for about 75 generations by which highest fitness show a plateau as a function of the generation number.*

## RESULTS AND DISCUSSION

**Evolutionary Screening of Sequence Space.** We conduct two evolutionary searches for each chain length – one for identifying optimal sequence that forms lowest radius of gyration and another one for identifying optimal sequence that forms highest radius of gyration. This strategy ensures that we sample data over the entire range of structures possible as a function of monomer sequence. Figure 2 shows the performance of the evolution algorithm for polymers of chain lengths N=20, 50, 100 and 150. For each case, we plot the objective function viz. the radius of gyration of the fittest candidate as a function of the number of candidates screened during the evolution process. For both minimization and maximization runs, there is a rapid progress at the beginning wherein the algorithm identifies better candidates in every generation. Subsequently, the search process reaches a saturation and the search is terminated when there is no improvement over more than 10 generations. The MD snapshots of the minimum and maximum radius of gyrations for a representative case of N=150 are shown in Figure 2c and

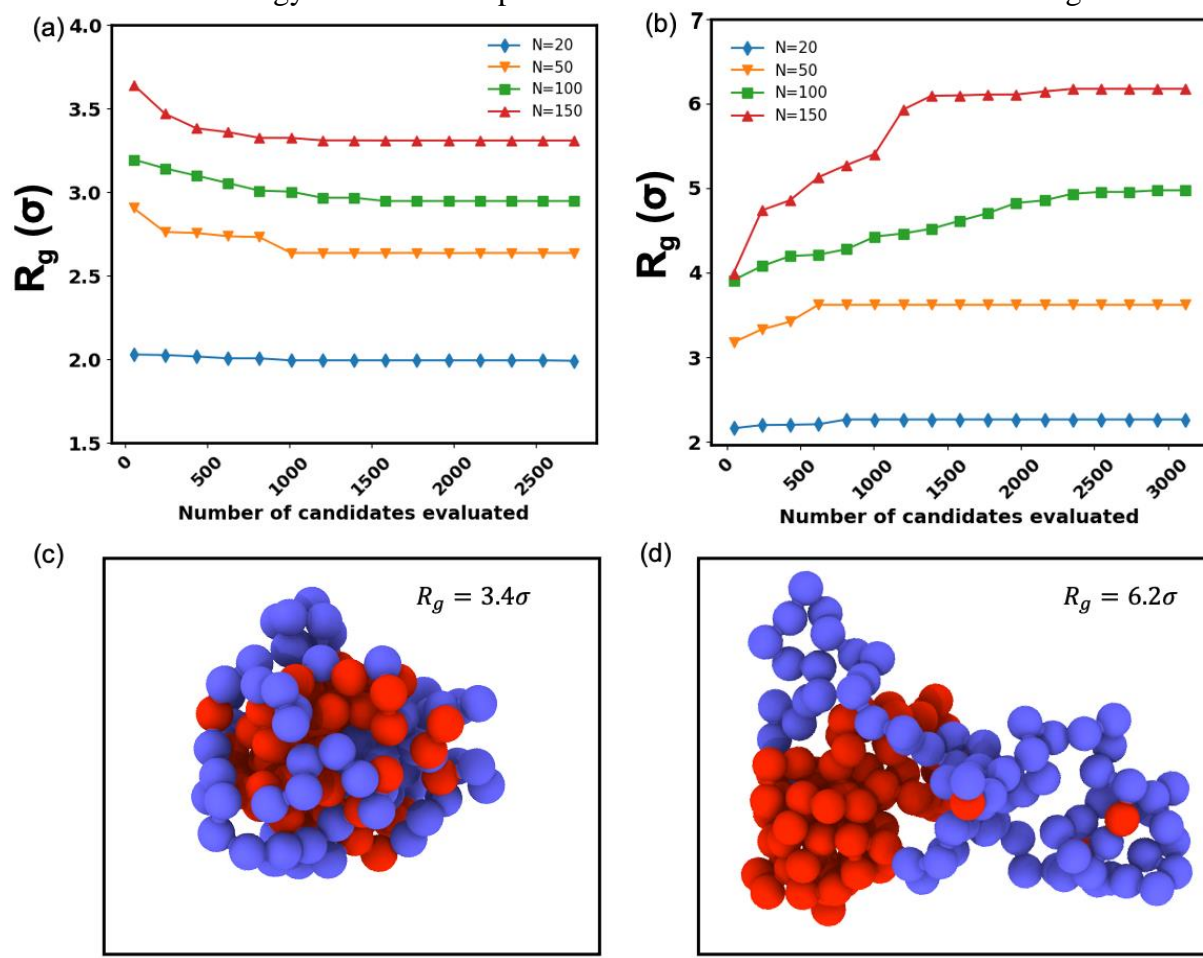

*Figure 2: The evolution of the materials design algorithm. The radius of gyration of the fittest candidate in a given generation is plotted as a function of the total number of candidate screened for minimization run and maximization run in (a) and (b), respectively. The 3D structure of polymer conformation for N=150 is shown for the two optimal sequences that correspond to lowest and highest radius of gyration in (c) and (d), respectively. The red and blue beads are A and B type moieties, respectively.*

2d, respectively. A list of all the optimized sequences, which are identified by our MD simulation based evolutionary algorithm (MD-EA) can be found in the supporting information (SI) along with their MD snapshots of equilibrium structures. For both the minimum radius of gyration and maximum radius of gyration, the sequence of A and B type monomers in the copolymer appears to be disordered or non-periodic. For a given polymer chain, a total of ~ 5000 sequences are evaluated during the two evolutionary computing runs. A cumulative total of 20000 sequences are studied using MD simulations within our evolutionary computing framework spanning polymer chain length from *N=20* to *N=150*. We separately run two MD simulations with pristine diblock copolymer and alternating block copolymer for all the four chain lengths, outside the evolutionary computing workflow. These two periodic sequences – diblock and alternating block copolymers are the two extreme cases of monomer distribution in a copolymer chain, and serve as a baseline to compare the properties of all other sequences. For the alternating block copolymer, the mean block length is $1\sigma$ and for the diblock copolymer the mean block length is $N\sigma/2$ for a polymer of chain length *N*. The mean block length of all other possible sequences falls within this two extreme values. We now compare the radius of gyration of the optimal candidates along with the diblock and alternating block copolymer chains in Figure 3 for all the chain lengths considered in this study. The data suggest that the range of $R_g$ possible in a copolymer increases with the polymer chain length. For very short chain like *N=20*, the impact of sequence on the radius of gyration of the chain is minimal. However, for *N=150*, The lowest and highest $R_g$s are $3.4\sigma$ and $6.2\sigma$, respectively. This is over more that 80% variation in radius of gyration as a function of sequence. It also clearly indicates that the range of $R_g$ will further expand rapidly for higher chain lengths. The $R_g$ of the EA maximized sequence is very close to that of a diblock polymer for all the cases as shown in Figure 3. For example, the EA maximized sequence for N=150 is 71A-1B-1A-35B-1A-19B-1A-19B-1A-1B, while that of a diblock copolymer is 75A-75B. The only difference between the EA maximized sequence and diblock sequence is that four A moieties are sparsely distributed in the B dominated half of the chain. Therefore, we infer that a copolymer, whose monomer distribution is close to that of a AB type diblock polymer, will have very high radius of gyration than any other distribution of moieties. On the other hand, visual analysis of the arrangement of A and B type moieties in the EA minimized sequence reveals that the polymer consists of multiple A segments and B segments. The lengths of these segments are not equal and there is no specific order of A and B segments in the polymer topology. For example, the sequence of As and Bs in a EA minimized copolymer of chain length *N=150* is 2B-2A-5B-11A-2B-1A-6B-2A-4B-3A-1B-1A-1B-1A-5B-3A-7B-6A-5B-1A-2B-4A-4B-1A-4B-1A-1B-3A-2B-4A-1B-3A-7B-4A-2B-1A-1B-1A-6B-10A-4B-1A-1B-1A-1B-2A-1B-8A-1B. Evidently, the polymer is made of segments ranging from block length $1\sigma$ to $11\sigma$. The combination of short segments – as short as $1\sigma$ and long segments – as long as $11\sigma$ whithout any specific order leads to the highest compact structure with lowest $R_g$. Interestingly, the $R_g$ of the EA minimized sequence differs significantly from that of an alternating block copolymer, which is $(AB)_{75}$ for *N=150*. For example, for N=150, the $R_g$ of the EA minimized sequence is $3.4\sigma$, while the $R_g$ for the alternating block copolymer is $4.5\sigma$. This is a reduction of more than 20% in $R_g$ value from that of an alternating block copolymer. Visual inspection of these EA-minimized structures reveals that these are protein-like aggregates. The core of these aggregates are formed by A type moieties, and the surface of these aggregates are formed predominantly by B type moieties.

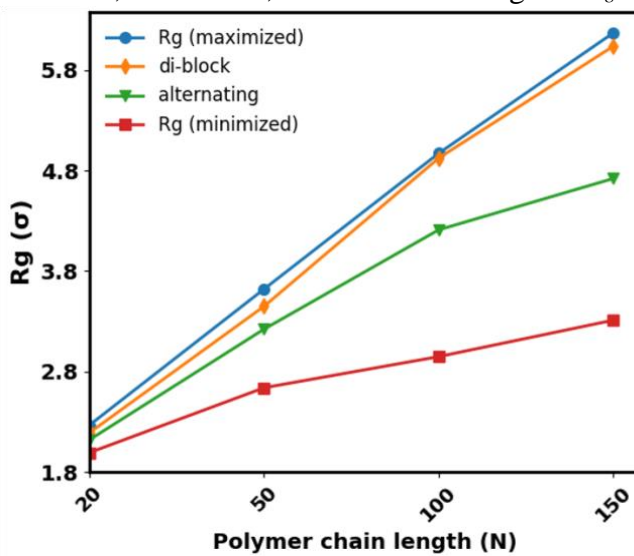

*Figure 3: Range of Single molecule structure. Radius of gyration of sequence specific copolymers as function of chain length is show for four different cases – diblock copolymer, alternating block copolymer and two evolutionary algorithm identified optimal sequences.*

**Sequence-Structure Correlations.** We now study the sequence-structure mapping of all the candidates. To understand this mapping of sequence space to the structure space, we calculate the fraction of sequences among the EA screened candidates, for a given chain length, that have very close $R_g$ values. Figure 4 represents the histogram of $R_g$ for two representative cases -

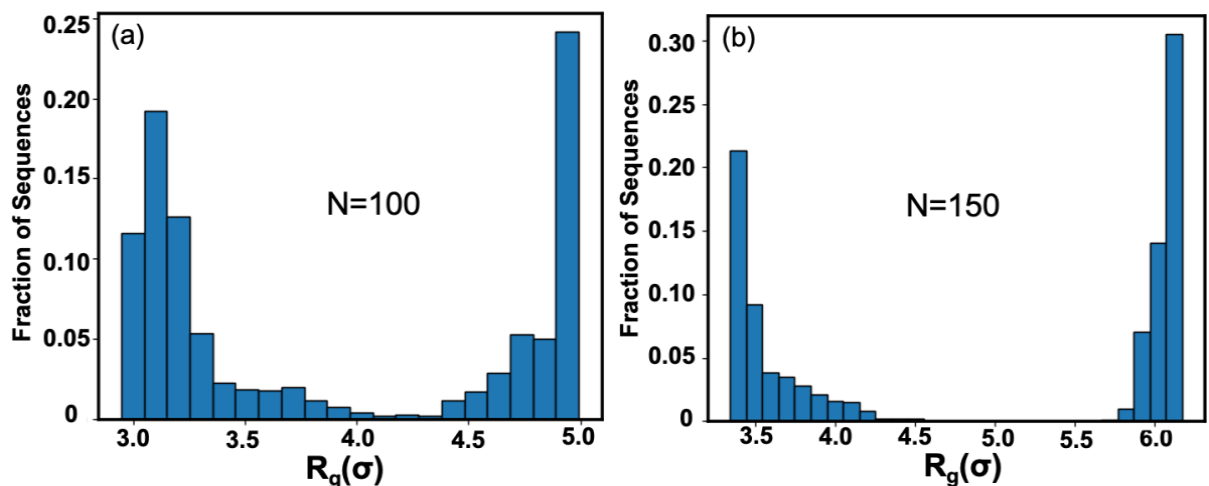

*Figure 4: Degeneracy of copolymer structures. Histogram of the radius of gyrations that are screened during the evolutionary search for (a) N=100 and (b) N=150. The total number of candidate structures used in histogram calculations are 5000 for both (a) and (b).*

$N=100$ and $N=150$. It indicates a broad distribution of $R_g$ possible in the model system as a function of sequence. It also clearly suggests many-to-one mapping between the sequence space and structure space wherein multiple candidates in the sequence space are connected to one point in the structure space. Now, the question is why so many different sequences forming similar structures? Is there any structural descriptor/feature that are common among these sequences that lead to same $R_g$? To further understand this similarity in the 3D structures of a copolymer chain as a function of its monomer sequence, we compute the mean block length of all the sequences. The mean block length of a copolymer with $n$ blocks can be written as $L_b = \sum_i l_i/n$, where $l_i$ is the length of the i$^{th}$ block. There have been many recent efforts to correlate copolymer properties such as thermal conductivity and surface tension with its mean block length.[28,29,38] Here, we aim to understand the correlation between mean block length and $R_g$ among all the sequences that are identified by the EA runs. We plot the radius of gyration of all the sequences screened during the evolutionary search as a function of their mean block length in Figure 5 for the same two representative cases $N=100$ and $N=150$. Although the exact sequences of A and B of all the candidate copolymers are different, but a subset of them have same mean block length. In fact, many of the chains, which show same $R_g$, have same mean block length $L_b$, in spite of different order of As and Bs in their architecture. However, Figure 5 also shows many exceptions, especially in the longer mean block length region. In the long mean block length region, many of the candidate sequences lead to different $R_g$s in spite of having same mean block length. Overall, the mean block length of a sequence does not show any strong correlation with the radius of gyration for the entire range of $R_g$ and $L_b$ that are captured in this study. However, there is a narrow range of mean block length for which $R_g$ broadly decreases as $L_b$ increases. On the other hand, copolymers with higher mean block length form very wide range of structures – both high and low $R_g$s. The perpendicular dotted line in Figure 5 qualitatively shows the cross-over of these two regions. The left side of the dotted line, we observe a qualitative inverse correlation between $L_b$ and $R_g$. On the right side of the dotted line, such correlation can't be drawn as the variation of $R_g$ is very large. We also collect a subset of sequences that are periodic for a given chain length and added them in Figure 5. As show by circular legends and connected by solid line in Figure 5, there is a modest correlation between their radius of gyration and mean block length. Considering only the periodic sequences possible for a given chain length, we observe that the radius of gyration decreases as a function of mean block length initially, and then it increases for the further increment in mean block length. The EA-minimization run yields large number of sequences that form compact structures with lower Rg than that an alternating block copolymers. These sequences are protein like sequences. The degree of "blocky-ness" in these MD-EA identified "protein-like' sequences are higher that a random copolymer or an alternating block copolymer. Unlike any random sequences, these protein-likes sequences exhibit long range correlations which can be described by Levy-flight statistics.[58]

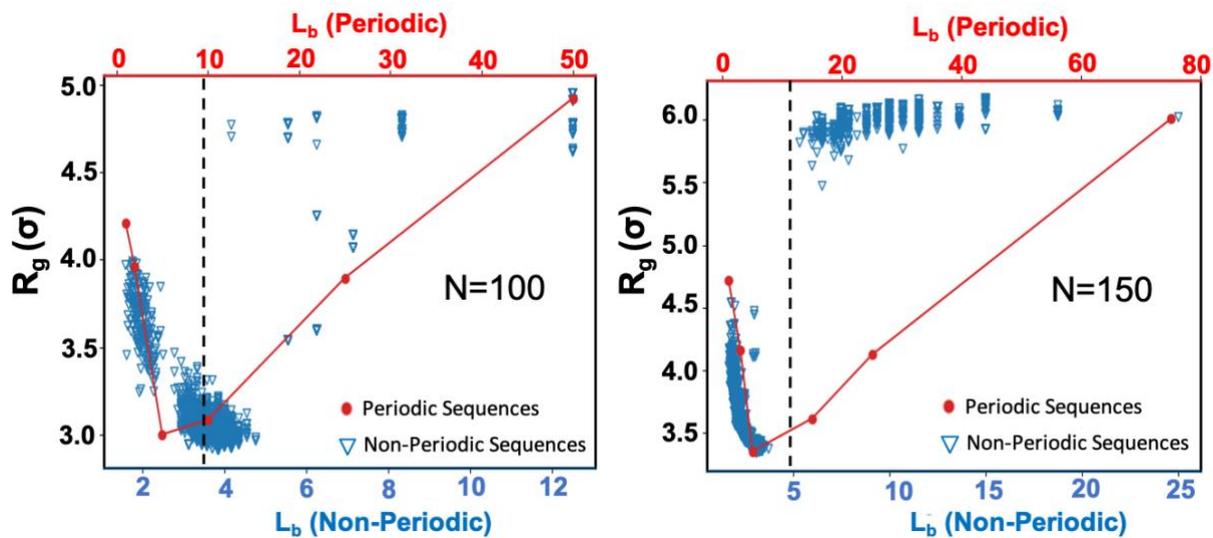

*Figure 5: $R_g$-mean block length correlation. Radius of gyrations of all the sequences screened during evolutionary computing are plotted as a function of their mean block length for chain lengths $N=100$ and $N=150$ in (a) and (b), respectively. The data points are connected by solid red line for periodic sequences as a guide to eyes. The bottom x-axis represents mean block lengths for non-periodic sequences while the top x-axis corresponds to the mean block length of the periodic sequences. The dotted perpendicular line separate two regions – The left side of the dotted line shows a modest $R_g$-$L_b$ correlation, while the right side of the line indicates a large variation of $R_g$ as a function of $L_b$.*

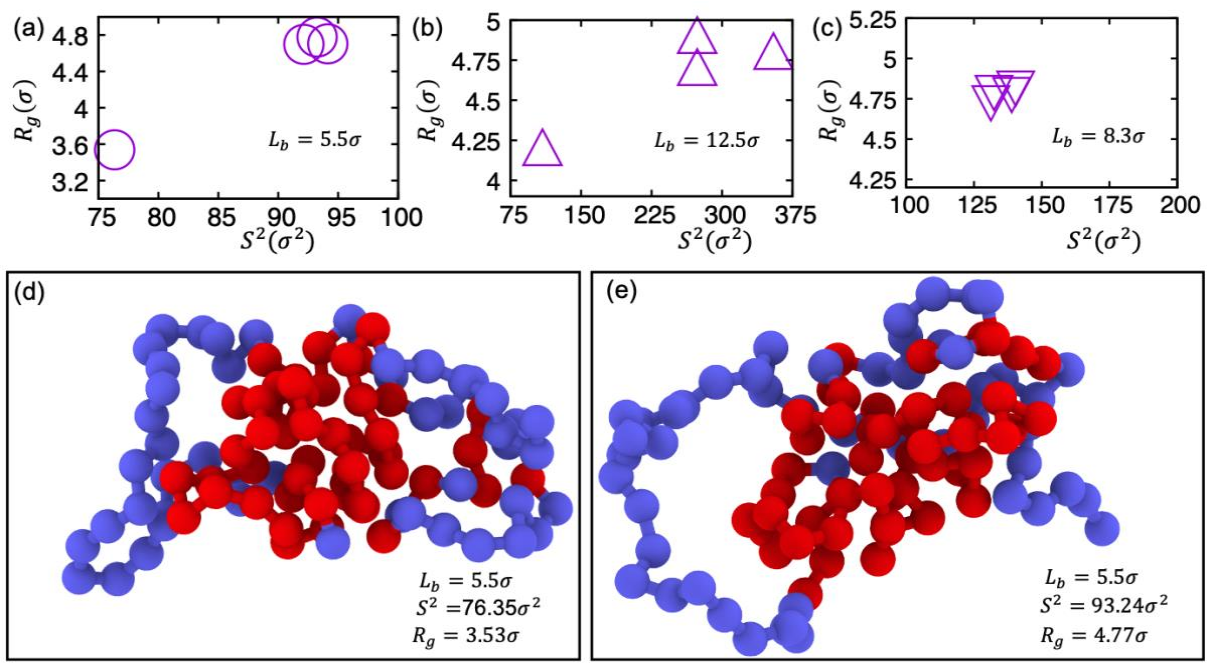

*Figure 6: Variance of polymer structure. The radius of gyration of a polymer chain is plotted against its variance of block length for a fixed mean block length in (a), (b) and (c) for $L_b = 5.5\sigma$, $L_b = 12.5\sigma$ and $L_b = 8.8\sigma$, respectively. In equilibrium MD snapshots are shown in (d) and (c) for $S_2 = 76.35\sigma^2$ and $S_2 = 93.24\sigma^2$, respectively for a fixed mean block length of $5.5\sigma$. The red and blue particles are A and B type moieties, respectively.*

Now, we intend to understand the variability of the radius of gyration among the polymer chains with fixed mean block length. We compute the variance in their block lengths to understand this Rg variability. The variance of block length of a polymer chain is defined as $S^2 = \sum (l_i - L_b)^2/(n-1)$. Here, $l_i$ is the length of i$^{th}$ block of a polymer chin with $n$ number of blocks. We select four polymer chains that have same mean block lengths however their radius of gyrations vary significantly, for the case of *N=100*. We aim to explain the variability of the radius of gyration as a consequence of the variances of block length in the polymer chain. Figure 6a and b presents two such cases where the radius of gyration is plotted against $S^2$ for $L_b$=*5.5σ* and *12.5σ*. For L$_b$=5.5σ and 12.5σ, there is clearly two regimes – once corresponding to low $R_g$ and the other corresponds to high $R_g$ as shown in Figure 6a and 6b, respectively. Among the four cases, one is in the lower $R_g$ regime while the remaining three are in the higher $R_g$ regime. The lower $R_g$ is seen when the block length variance of the polymer chain is low. On the other hand when the block length variance is high, we observer higher radius of gyration. Further, we choose a case where the mean block length is same and the block length variances are reasonably close. As shown in Figure 6c, the $R_g$ values of all the four candidates are similar. Thus, the combination of mean block length and block length variance provide a plausible explanation for two questions – 1. It demystifies the variability of $R_g$ among the candidates, particularly in the long $L_b$ region. These are the sequences that have same $L_b$ but significantly different $S^2$. 2. It explains why many different sequences of a copolymer lead to a single $R_g$, particularly in the short $L_b$ region. These are the sequences that have same $L_b$ as well as same $S^2$ in spite of varying the order of A and B moieties in the polymer architecture.

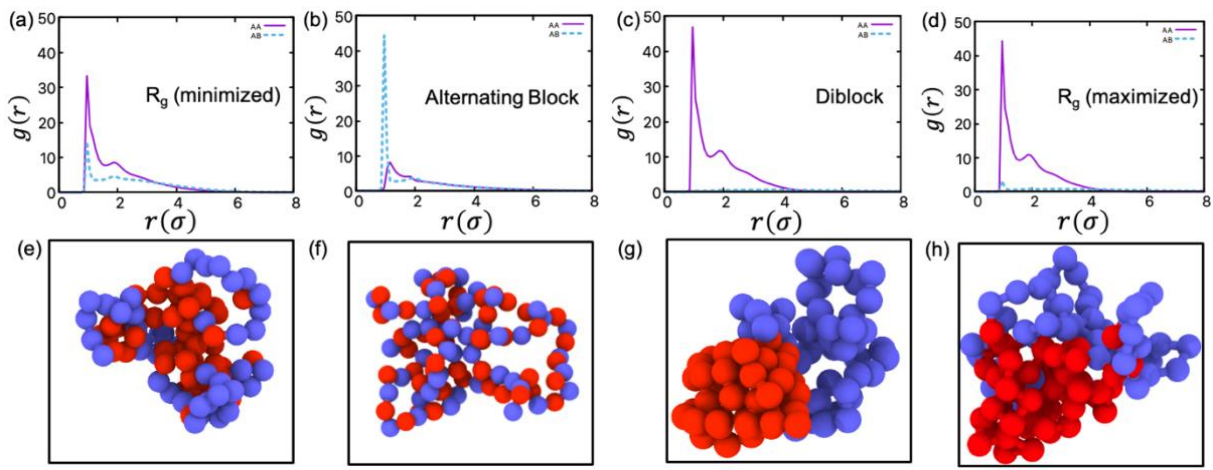

*Figure 7: Radial distribution of monomers in a single molecule. The radial distribution function of A and B type moieties around an A type moiety is shown for Rg-minimized, alternating block, diblock and Rg-maximized sequences in (a), (b), (c) and (d), respectively. The corresponding in equilibrium molecular snapshots are shown in (e), (f), (g) and (h). Here, red and blue bead represent A and B type moieties, respectively. The red and blue particles are A and B type moieties respectively.*

**Sequence Defined Packing of Single Molecule:** In this section, we focus on local structure of a copolymer and study the impact of sequence on the single molecular packing for a representative case of *N=100*. We calculate radial distribution function which is defined as the local density of a specific moiety around a given moiety as a function of separation distance between them. Figure 7a, b, c and d represent radial distribution function for Rg-minimized sequence, alternating block copolymer, diblock copolymer and Rg-maximized sequence, respectively. The MD snapshots of the corresponding four molecular conformations are shown in Figure 7e, f, g and h for EA-minimized sequence, alternating block copolymer, diblock copolymer and EA-maximized sequence, respectively. For each of these cases, we compute two radial distribution functions – one for the A-A pair and the second one is for the A-B pair of moieties. For the EA minimized sequence, the local density of A type moiety around an A type moiety $g(r_{AA})$ is higher than the local density of B type moiety around an A type moiety, $g(r_{AB})$, as shown by the first peak heights of the two functions in Figure 7a. However, for an alternating block copolymer the first peak of $g(r_{AB})$ is significantly higher than that of the $g(r_{AA})$. As the sequence of the alternating block is (AB)$_{50}$, which is made of AB dimers only, the distribution of B around A is higher than that of A around an A. On the other hand, the EA minimized sequence is 3B-6A-3B-4A-8A-2A-1B-3A-1B-4A-5B-1A-4B-8A-1B-1A-3B-1A-7B-10A-6B-1A-1B-1A-4B-4A-1B-1A-2B-3A, which contains AB dimers as well as large number of pure A and B strings with varying size. A balanced combination of $g(r_{AB})$ and $g(r_{AA})$ lead to a very compact structure with lowest $R_g$ in the EA minimized sequence. For the diblock copolymer, the AA pair distribution function shows highest first peak intensity and AB pair distribution function exhibits lowest pick intensity, as shown in Figure 7c. This leads to a phase separation of the two blocks as visually seen in Figure 7g. The distribution function for EA-maximized sequence is very similar to that of the dibock case with slightly higher density for the AB pair. This is due to the fact that EA maximized sequence (47A-25B-1A-20B-1A-4B-1A-1B) has 3 A-type moieties that are sparsely placed in the B dominant half of the chain. This is also in line with the fact that the $R_g$ of a diblock copolymer is very close to that of the EA maximized sequence. A strong $g(r_{AA})$ and a very weak $g(r_{AB})$ lead to extended structure with highest $R_g$ in the EA maximized sequence. The packing of the polymer can be further understood by analyzing the energy of the systems. Figure 8 shows the in equilibrium Lennard-Jones energy of the system for the four cases. As the diblock forms a complete phase separation of A and B type moieties, it has maximum favourable interaction (A-A interactions), the system has lowest pair energy as shown in Figure 8. For the alternatiing block copolymer, A-B pair density is strongest and it accounts for maximum unfavourable interactions (A-B interactions). Therefore, the energy of the alternating block copolymer system is highest. For the two EA optimized sequences, the pair energy is in between that of a diblcok and an alternating block copolymer. For the EA maximized sequence, the presence of three BAB trimers in the B dominant region of the chain leads to its slightly higher energy in comparison to that of the diblock copolymer. For the case of EA minimized sequence, the number of ABA trimers are higher than that of a diblock but lower than that of the alternating block copolymer. This leads to the total pair energy higher than that of the diblock and lower than that of the alternating block copolymers.

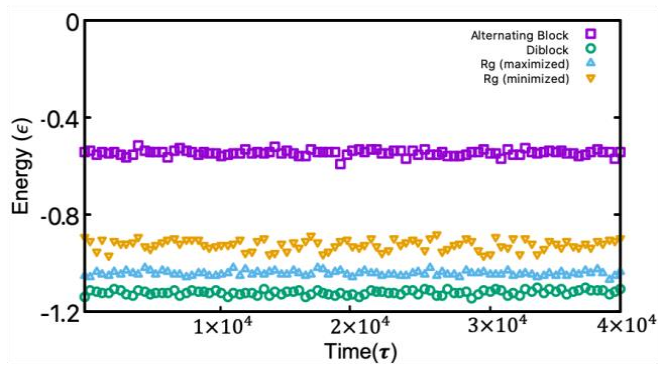

*Figure 8: Pair interaction energy of a copolymer. The per particle Lennard-Jones interaction energy of four systems – alternating block copolymer, diblock copolymer, EA maximized sequence and EA minimized sequence during production runs are shown as a function of MD time.*

**Sequence Specific Packing of Copolymers in a Material.** We further study the impact of sequence in a bulk polymeric material via isothermal isobaric molecular dynamics simulations. Specifically, packing of the EA-optimized sequences along with alternating and diblock copolymers in a bulk system are tested for a representative case of polymer chain length

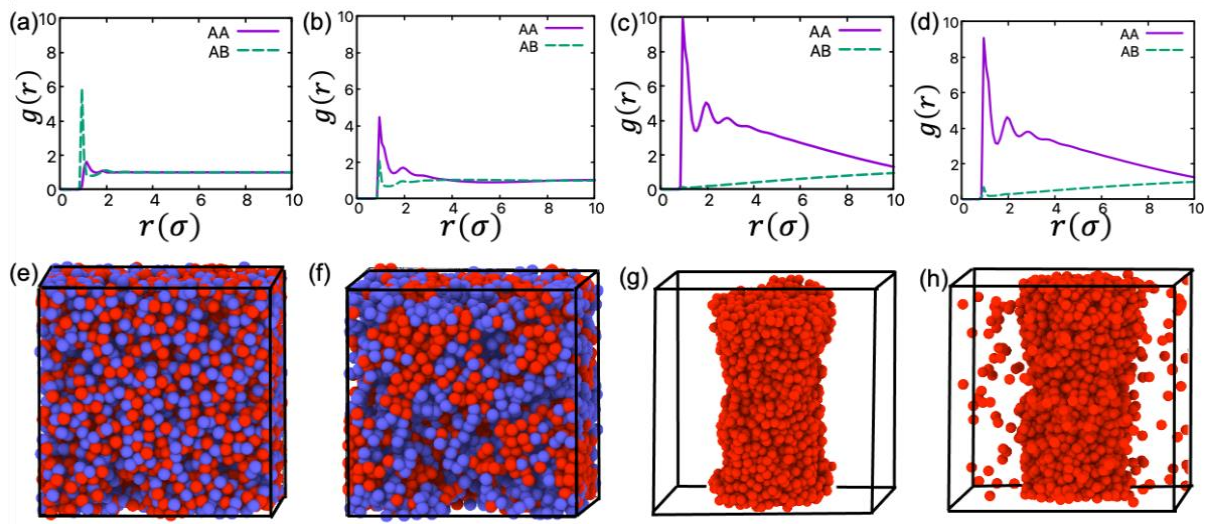

*Figure 9: Copolymer packing in a bulk system. The radial distribution function for alternating block copolymer, EA minimized sequence, diblock copolymer and EA maximized sequence are shown in (a), (b), (c) and (d), respectively. For all the cases two radial distribution functions are shown – one for AA pair and the second one for AB pair, as mentioned in the legends. The in equilibrium molecular conformations are shown in (e), (f), (g) and (h) for alternating block copolymer, EA minimized sequence, diblock copolymer, EA maximized sequence, respectively. The read and bule colours represent A and B type moieties, respectively. The B type moieties are not shown in (g) and (f) for better visual inspection of the structures.*

*N=100*. We note that these sequences are optimized for solution. It is quite possible that the optimal sequences for bulk system are different. Here, our intention is to test whether it is possible to develop sequence specific copolymer that fall outside the known range of density of a bulk copolymer system defined by the two extremes – diblock copolymer and alternating block copolymer. Evidently, our simulation shows that the equilibrium number densities are $0.48\sigma^{-3}$, $0.52\sigma^{-3}$, $0.34\sigma^{-3}$ and $0.33\sigma^{-3}$ for alternating block copolymer, EA-minimized sequence, diblock copolymer, EA-maximized sequence, respectively. First of all, this suggests more than 50% change in the density of the system by changing the copolymer sequence from the EA-maximized sequence to the EA-minimized sequence. Secondly, the density of our EA-minimized sequence is greater than that of an alternating block copolymer. In principle, addressing the question of highest possible density in a copolymer system requires conducting optimization calculation with bulk density as the fitness function. This is not within the scope of the current study. Nonetheless, our data clearly indicates that the range of possible density in a bulk copolymer system can be expended by monomer level sequence control. We further, analyze the arrangement of A and B type moieties for these four systems. Figure 9 show the radial distribution function for all the cases along with MD snaps shots. In these bulk systems, the distribution of B type moieties around an A type moiety is similar to that observed for single molecule packing. In the alternating block copolymer system, the height of the first peak of $g(r_{AB})$ is significantly greater than that of the $g(r_{AA})$ as the system consists of AB dimers, which is very clear from Figure 9a. On the other hand, the EA-minimized sequence, the $g(r_{AA})$ peak is higher than that of the $g(r_{AB})$, as shown in Figure 9b. There is a significant aggregation of A type moieties and B type moieties separately in the system. These local domains of A moieties and B moieties form a percolating gyroid-like network structure as shown in Figure 9f. It is also significantly different form that of the alternating block copolymer where A and B type moieties are uniformly distributed in the system as shown in Figure 9e. Moving from EA-minimized sequence to a diblock copolymer, we see a complete phase separation. In the diblock system, $g(r_{AA})$ shows highest first peak, and the $g(r_{AB})$ peak height is minimal as shown in Figure 9c. This indicates a phase separation between A and B type moieties. The $g(r_{AA})$ peak height slightly reduces when the diblcok polymer is replaced by the EA-maximized sequence as depicted in Figure 9d. This reduction leads to a sparse dispersion of A type moieties. The bulk density of a diblock copolymer system is close to that of the EA maximized sequence. But their microstructures have noticeable differences. As shown in Figure 9g, the diblock copolymer system forms a perfect cylindrical micelle structure wherein the A type moieties form a cylinder within the simulation box, and the remaining volume fraction is occupied by B type moieties. However, for the EA-maximized sequence, in addition to a cylindrical structure of A type moieties, few A moieties are sparsely distributed in the B rich domain of the simulation box as shown in Figure 9h. Therefore, this analysis suggests a wide range of new structures and functions along with well know spherical micelles, cylindrical micelles, lamellar and gyroids structures[3,59–61] can be leveraged by engineering monomer level sequence of a copolymer.

**CONCLUSIONS**

Characterizing the impact of sequence on the 3D shape of a macromolecule is the key to understand the functions of many biological systems including protein and DNA. Moreover, an improved understanding of the impact of sequence on a molecular topology will enable a more effective control of structure-property relations in synthetic polymeric materials and open a viable pathway to tailor materials properties that are essential for miniaturized, more efficient and durable next generation devices. The central barrier in designing such materials is that the exploration of sequence space of a copolymer is challenging due to the astronomically large number of possible combinations. Therefore, efficiently generating and analyzing the big data of copolymer sequences and structures have been one of the major challenge in polymer informatics. Recent progress in machine learning and advanced optimization methods along with coarse-grained MD simulations can play a vital role for this purpose. Herein, we use one such approach viz., MD simulation based evolutionary algorithm to design sequence specific polymer and study sequence-structure relationship of a model AB type linear copolymer with equal composition. Within this framework, starting from a random copolymer, we repetitively employ evolutionary operations – selection, mutation and crossovers to identify new sequences and conduct CGMD simulations to measure their equilibrium structures. During this CGMD simulation based evolutionary computing, we determine in-equilibrium structures of more than 20000 sequences of the model copolymer, spanning across different chain lengths. The EA search spans across the two extremal points of highest and lowest radius of gyrations for all the chain lengths. We find that the optimal sequences for the lowest Rg and highest $R_g$ are non-periodic and non-intuitive for all the chain lengths. However, we note that the sequence of our EA-identified copolymer with highest Rg is very close to that of a diblock copolymer, where half of the chain is predominantly consists of A type moieties, and the remaining half of the chain is predominantly consists of B type moieties. The EA-identified copolymers with lowest

Rg yield protein-like aggregates. Identifying protein-like synthetic copolymer is always a challenging design task. Here, starting from a random sequence of AB type copolymer with equal composition and setting Rg as the fitness of an evolutionary search, we are able to track specific sequences that fold like a globular protein. We also demonstrate that the structures of a copolymer with non-periodic sequences deviate significantly from that of periodic sequences. The mean block length and the block length variance of a copolymer together provide a plausible explanation for the variation in $R_g$ as a function of sequence. As the current study is restricted to AB type copolymer with equal proportion of A and B type moieties, future studies are required towards understanding a universal sequence-structure correlations of copolymers with varying ratios of moieties as well as the presence of side chains and other possible molecular architectures. The work illustrates the non-periodic sequences are very important and monomer-to-monomer sequence control of a random copolymer can expand the range of 3D structures possible in a copolymer system. We also test the packing of EA-identified solution phase optimal sequences in a bulk environment. It suggests that packing density of a bulk system can be enhanced by monomer level sequence specific control. It would be of further interest to optimize the density of the bulk system as a function of sequence. Over all, current study demonstrates the effectiveness of evolutionary computing for sequence engineering of macromolecules for target structures and functionalities, and reports new insights on the sequence-Rg correlation of copolymers.

**SUPPORTING INFORMATION**

The MD-EA identified optimal sequences of all four polymer chains and the single molecule structures of the optimal sequences.

**NOTES**

The authors declare no competing financial interest.

**ACKNOWLEDGMENT**

The work is made possible by financial support from SERB, DST, Gov of India through a start-up research grant (SRG/2020/001045) and National Supercomputing Mission's research grant (DST/NSM/R&D_HPC_Applications/2021/40). TKP acknowledges ICSR, IIT Madras for initiation, seed and exploratory research grants. This research used resources of the Argonne Leadership Computing Facility, which is a DOE Office of Science User Facility supported under Contract DE-AC02-06CH11357. We also used computational facility of the Center for Nanoscience Materials. Use of the Center for Nanoscale Materials, an Office of Science user facility, was supported by the U.S. Department of Energy, Office of Science, Office of Basic Energy Sciences, under Contract No. DE-AC02-06CH11357.We acknowledge the use of the computing resources at HPCE, IIT Madras.